# Synthesis, structure and magnetic properties of Fe@Pt core-shell nanoparticles

K. Pisane, S. Singh, and M. S. Seehra*

Department of Physics & Astronomy, West Virginia University, Morgantown, WV 26506-6315

*Address correspondence to this author.  Email: mseehra@wvu.edu

ABSTRACT

Structural and magnetic properties of Fe@Pt core-shell nanostructure prepared by a sequential reduction process are reported. Transmission electron microscopy (TEM) shows nearly spherical particles fitting a lognormal size distribution with $D_o$= 3.0 nm and distribution width $\lambda_D$= 0.31. In x-ray diffraction, Bragg lines due to Pt shell only are clearly identified with line-widths yielding crystallite size =3.1 nm. Measurements of magnetization M vs. T (2 K - 350 K) in magnetic fields up to 90 kOe show a blocking temperature $T_B$ = 13 K below which hysteresis loops are observed with coercivity $H_C$ increasing with decreasing T reaching $H_C$ = 750 Oe at 2 K. Temperature dependence of the ac susceptibilities at frequencies $f_m$ = 10 Hz to 5 kHz is measured to determine the change in $T_B$ with $f_m$ using Vogel-Fulcher law. This analysis shows the presence of significant interparticle interaction, the Neel-Brown relaxation frequency $f_o$ = 5.3 x $10^{10}$ Hz and anisotropy constant $K_a$ =3.6 x$10^6$ ergs/ cm$^3$. A fit of the M vs. H data up to H = 90 kOe for T > $T_B$ to the modified Langevin function taking particle size distribution into account yields magnetic moment per particle consistent with the proposed core-shell structure; Fe core of 2.2 nm diameter and Pt shell of 0.4 nm thickness.

1. **Introduction**

Rare and precious metals Pt, Pd and Rh are currently used in a variety of key catalytic reactions such as automobile three-way convertors [1]. So, reducing their use is a national priority according to a 2013 report from the U. S. Department of Energy. An important approach in this regard is the use of bimetallic catalysts of the precious metals with iron-group transition metals, particularly with core-shell morphology with the shell made of a precious metal. Core-shell nanoparticles (NPs) are a special class of nanostructured materials whose properties depend not only on the constituents but also on the core-shell volume ratio [2,3]. In this regard, here we present our results on the synthesis, structural characterization and magnetic investigations of a 3 nm Fe@Pt core-shell nanostructure. It is noted that the obtained structure is not FePt alloy, a system whose properties have attracted a great deal of attention in recent years [4-8]. The relevant studies in connection with the results presented here are those reported on the core-shell NPs of CoO@Pt [9], Fe@Au [10], Fe@Ag [11], Co@Pt [12], Fe@$\gamma$-$Fe_2O_3$ [13], multi-shell Fe@$Fe_3O_4$/$Fe_2O_3$@FePt@Pt nanostructure [14], and a number of studies on Fe NPs of various sizes [15-19]. Details of our procedure for synthesizing the Fe@Pt NPs, their structural characterization by a number of analytical techniques and measurements, and interpretation of their magnetic properties are given below.

2. **Synthesis and Structural Characterization**

Fe@Pt core-shell nanoparticles (NPs) were synthesized by a sequential reduction process adopting and modifying the procedure described by Alayoglu et al. for synthesizing Ru@Pt NPs [20] and by Zhou et al. for synthesizing Cu@Pt NPs [21]. The precursors used were iron acetylacetonate ($C_{15}H_{21}FeO_6$, F.W. =353), ethylene glycol ($C_2H_6O_2$, F.W. =62.07), platinum chloride (PtCl$_4$, F.W. =336.9) and polyvinylpyrrolidone (PVP), all obtained from Alfa Aesar Inc. and used as received. The iron core was first synthesized by the reduction of 40 mg of acetylacetonate with 20 mg of ethylene glycol and 25 mg of the stabilizing agent PVP. These precursors were mixed and the solution was refluxed at 160-180 $^0$C for one hour. After cooling down to room temperature, 45 mg of PtCl$_4$ was added to the suspension followed by additional refluxing for 3 hours under air atmosphere resulting in a black colloidal solution. After cooling to room temperature, 25 mL of hexane was added to precipitate the nanoparticles. The collected precipitates were subsequently washed in ethanol, acetone and hexane. Finally, the precipitated sample was annealed at 200 $^0$C for 2 hours in ultra-high purity flowing N$_2$ gas.

Morphology and size of the NPs were investigated by transmission electron microscopy (TEM) using a JEOL JEM-2100 system. In Fig. 1, the measured diameters of the nearly spherical particles using ImageJ software are fit to a lognormal distribution yielding $D_o$ =3.0 nm and distribution width $\lambda_D$=0.31, which in turn yield average $<D>$ =3.1 nm and standard deviation $\sigma$ =1.0 nm. In Fig. 1c, TEM micrographs are shown along with the observed lattice fringes (Fig. 1b) corresponding to d (111) = 0.226 nm for Pt.

The x-ray diffraction (XRD) pattern of the synthesized sample using CuK$_\alpha$ source ($\lambda$=0.15418 nm) is shown in Fig. 2 with the expected line positions marked for Pt, FCC FePt, and α-Fe. The weak broad line near $2\theta \approx 22^0$ corresponds to polycrystalline PVP [22] used as a stabilizer in the synthesis. The observed lines are indexed for FCC Pt although the expected line positions from FCC FePt alloy are nearby since the lattice constant of Pt (3.92 Å) is only slightly larger than that of FCC FePt (3.84 Å). There is no clear evidence for lines due to α-Fe. This is similar to the reported cases of XRD in core-shell NPs of CoO@Pt [9], Fe@Au [10], and Co@Pt [12], and the complex Fe@Pt nanostructures [14], where XRD lines due to the cores were also not observed. Several reasons for this are possible: First, the exposure of the core to x-rays is blocked, at least partially, by the shell; second, the atomic scattering factors of Pt and Au are about 3 times larger than that of Fe and Co making the Bragg lines from Pt and Au that much more intense; and third, the core may be amorphous as suggested by Zhang et al. [23] in their studies on the carbon supported Fe@Pt NPs. Using the widths of the major lines in Fig. 2 and the Scherrer relation, the average crystallite size $\langle D \rangle$ =3.05 (0.20) nm is determined in good agreement with the TEM results.

Considering the amounts of precursors used in the synthesis of the Fe@Pt NPs of 3 nm average diameter, we have calculated the core diameter = 2.2 nm, Pt shell thickness= 0.4 nm, the atomic ratio Fe/Pt= 0.85, and the weight of Fe = 19.6 % of the total weight. These calculations were done without considering PVP. Results from the thermo-gravimetric analysis (TGA) of PVP and the Fe@Pt NPs (see Fig. 3) show that nearly 73 % of the weight of the sample is due to PVP despite extensive washing with acetone, ethanol and hexane. Lines due to PVP are detected in the IR spectroscopy of the sample (results not shown here) suggesting coating of the NPs with PVP. These results show that the weight of Fe in the Fe@Pt nanostructure is only 5.4 % of the sample weight. This is used in the analysis of the magnetic data since the measured magnetization is due mainly to Fe, Pt being only weakly paramagnetic and PVP being weakly diamagnetic.

3. **Results from Magnetic Measurements**

All the magnetic measurements were done using standard procedures applied to the PPMS magnetometer by Quantum Design Inc. and the magnetization plotted is in units of emu/g-Fe using the 5.4 wt. % of Fe. In Fig. 4a, the temperature dependence of magnetization M vs.

temperature T measured in H=100 Oe for the traditional ZFC (zero field-cooled) and FC (field-cooled) cases shows M(ZFC) peaking at the blocking temperature $T_B$ = 13.0 K although the bifurcation between M(FC) and M(ZFC) begins to occur at about 18 K. This difference is due to the particle size distribution since higher temperatures are required to unblock the larger particles. Fig. 4b shows the plot of M vs. H at 2 K measured up to 90 kOe with the expanded view of the low field region shown in Fig. 5b. The loop is symmetric with coercivity $H_c \approx$ 750 Oe at 2 K and no indication of the saturation of M even at the maximum H=90 kOe. For comparison, for 3 nm NPs of the FCC FePt alloy, $H_C$ for T < $T_B$ was reported to be practically zero [7]. In Fig. 5a, $H_C$ for the Fe@Pt NPs is strongly temperature dependent but it becomes zero only at T > $T_B$ = 13 K because of contributions for larger blocked particles. In Fig. 5b, the differences in the low field behavior of the ZFC sample and the FC@ 15 kOe sample is shown, indicating a slight increase in $H_C$ for the FC sample. However, any loop asymmetry or exchange bias is less than 20 Oe, well within the experimental uncertainties.

The data of M vs. (H / T) for applied H up to 90 kOe for T= 50, 100, 150, 200, 250, and 300 K are shown in Fig.6a. Nonlinearity of M vs. H/T is evident and these data are used later to determine magnetic moment $\mu_p$ per particle. For determining the relaxation rate and the strength of interparticle interaction in the system, temperature dependence of the ac susceptibilities $\chi'$ and $\chi''$ were measured from 2 K to 20 K at frequencies $f_m$ from 10 Hz to 5 kHz using $H_{ac}$ = 10 Oe and $H_{dc}$ = 0 Oe (see Fig. 7). Theoretically, the magnitude of the $\chi''$ peaks at $T_B$ and $\chi''=C\partial(T\chi')/\partial T$ [24,25]. In Fig. 7, the data of $\chi''$ vs. T have considerably larger noise partly because the magnitudes of $\chi''$ are an order of magnitude smaller than those of $\chi'$. Since changes in the peak position of $\chi''$ to determine changes in $T_B$ with change in $f_m$ could not be determined with sufficient accuracy, the peak positions of $\partial(T\chi')/\partial T$ were used in the analysis given below.

4. **Analysis and Discussion**

From magnetic studies in a number of interacting nanoparticle systems [24-27], the variation of $T_B$ with change in $f_m$ follows the Vogel-Fulcher Eq.,

$$T_B = T_o + [T_a/\ln(f_o/f_m)] \quad (1).$$

Here, $f_o$ is the attempt frequency of the Neel-Brown relaxation, $T_o$ represents the strength of the interparticle interaction (IPI), and $T_a = K_a V/k$ with $K_a$ being the anisotropy constant, V the volume of the NPs, and k the Boltzmann constant. For $T_o = 0$ K, Eq. (2) reduces to the Eq. for Neel-Brown relaxation. The quantity $\Phi = \Delta T_B/T_B \Delta \log_{10} f_m$, where $\Delta T_B$ is the change in $T_B$ with change in $\log_{10} f_m$, also provides a good measure of the strength of the IPI. For $\Phi > 0.13$, IPI is negligible, $0.005 < \Phi < 0.05$ for spin glasses and for $0.05 < \Phi < 0.13$, IPI is present with its strength increasing with decreasing value of $\Phi$ [28,29]. The data in Fig. 7 yields $\Phi = 0.09$ suggesting significant IPI and hence $T_B > T_o > 0$ in Eq. (1). Using $T_o$ as a variable, the plot of $1/(T_B - T_o)$ vs. $\ln f_m$ was tried and the best fit is shown in Fig. 8 for $T_o = 5$ K yielding $T_a = 144(9)$ K and $f_o = 5.3 \times 10^{10}$ Hz. This magnitude of $f_o$ compares fairly well with $f_o = 1.8 \times 10^{10}$ Hz reported for Ni NPs dispersed in $SiO_2$ [25]. Using $T_a = 144$ K and particle diameter D = 2.2 nm for the Fe-core, yields $K_a = 3.6 \times 10^6$ ergs/cm$^3$ as the anisotropy constant, considerably larger than $K_a = 4.2 \times 10^5$ ergs/cm$^3$ for bulk Fe. This value of $K_a$ in the Fe@Pt NPs agrees well with $K_a \approx 2.5 \times 10^6$ ergs/cm$^3$ reported in Ref. 17 for Fe NPs. Recently, for the 2.3 nm surfactant coated Fe NPs, Monson et al. [16] reported $\Phi = 0.08$, $f_o = 3.3 \times 10^{11}$ Hz, $K_a = 1.9 \times 10^7$ ergs/cm$^3$ and $T_B = 16$ K. The major conclusions from this analysis are the evidence for the presence of significant IPI in the Fe@Pt NPs and an order of magnitude enhancement in $K_a$ vis-à-vis bulk Fe. The large coercivity $H_C$ (Fig. 6) is understandable in terms of enhanced $K_a$ because of the proportionality of $H_C$ and $K_a$.

We next consider the interpretation of the M vs. H data taken at several temperatures above $T_B$ (Fig.6b). Such data are often fitted to the modified Langevin function [30]:

$$M = M_o \mathcal{L}(\mu_p H/kT) + \chi_a H \quad (2)$$

where $\mathcal{L}(x) = \coth x - (1/x)$ and $\mu_p$ is the magnetic moment per particle. From Fig. 6a, it is evident that the data at the higher temperatures do not quite scale with H/T. This is due to the particle size distribution and the fact that, for the lower H/T values, the dominant contribution to M comes from only the larger particles [31]. Since the data at 50 K (100 K) reflect contributions from wider size distributions, we have fitted this data to Eq. (2) yielding $\mu_p = 617\ \mu_B$ (780 $\mu_B$), $\chi_a = 3.65$ (2.20), in units of $10^{-4}$ emu/g-Oe and $M_o = 7.04$ emu/g (4.21 emu/g). Considering the BCC unit cell of Fe with lattice constant = 0.287 nm and two atoms/unit cell, the number of atoms per particle with diameter D = 2.2 nm (2.0 nm) is calculated to be 472 (354). Using 2.22 $\mu_B$ per Fe atom for bulk Fe yields $\mu_p = 1047\ \mu_B$ (786 $\mu_B$) assuming complete alignment of all the moments. Thus magnitude of $\mu_p$ determined from the fit to Eq. (2) is consistent with the core diameter of about 2.0 nm.

For a system with a size distribution (Fig. 1a), it is appropriate to expect a distribution of moments. Therefore, we have also fitted the M vs. H data to the following lognormal distribution of the magnetic moments [32]:

$$M = \frac{N}{s\sqrt{2\pi}} \int_0^\infty \mathcal{L}(\mu H/k_B T) \exp\left\{\frac{-[\ln(\frac{\mu}{\mu_o})]^2}{2s^2}\right\} d\mu + \chi_a H \quad (3)$$

where $\mu_o$ is the median value of $\mu_p$ and s describes the width of the distribution. Fitting the data at 50 K (100 K) to Eq. (3) yields N = 1.19 x $10^{18}$ $g^{-1}$ (4.88 x $10^{17}$ $g^{-1}$), $\chi_a = 3.68$ x $10^{-4}$ emu/g-Oe (2.25 x $10^{-4}$ emu/g-Oe), $\mu_o = 587\ \mu_B$ (630 $\mu_B$) and s = 0.35 (0.73). These values yield $\langle\mu\rangle = 624\ \mu_B$ (822 $\mu_B$) with $\sigma = 225\ \mu_B$ (690 $\mu_B$), and $M_o = N\langle\mu\rangle = 6.89$ emu/g (3.72 emu/g). These results are consistent with those obtained from the fit to Eq. (2) for the M vs. H data at 50 K (100 K). As noted earlier, M vs. H data at the higher temperatures are dominated by the contributions from the larger particles only and so are not representative of the size distribution shown in Fig. 1.

The non-saturation of M vs. H even H up to 90 kOe (see Fig. 4b) is quite different from the observation in bulk Fe. Also, the magnitude of M ~ 70 emu/g-Fe at 90 kOe for the Fe@Pt NPs is considerably smaller than $M_S$ ~ 215 emu/g reported for bulk Fe. Both the non-saturation and the lower magnitude of $M_S$ likely result from canting of the spins especially at the surface for this 2.2 nm Fe core @Pt NPs. For comparison, other reported values of $M_S$ in Fe NPs are: $M_S$ ~ 40 emu/g at 10 K for 12 nm Fe@Au NPs [10]; $M_S$ ~ 179 emu/g at 4 K for the 4.6 nm Fe@Pt NPs with $T_B$ ~ 40 K and $H_C$ ~ 300 Oe at 4 K [14]; and $M_S$ ~ 210 emu/g at 5 K for 5 nm surfactant coated Fe NPs with saturation occurring above 30 kOe [16]. Thus, the size of the Fe core is the likely controlling factor for determining $M_S$.

## 5. Conclusions

Structural and magnetic characterization of the Fe@Pt core-shell NPs with Fe core diameter ~ 2.2 nm presented here show a $T_B$ = 13 K below which $H_C$ increasing with decreasing T with $H_C$ ~ 750 Oe at 2 K is observed. This relatively large value of $H_C$ and the non-saturation of the magnetization even at 90 K likely result from the smaller size of the core which in turn produces large fractions of the canted spins at the surface of the core. The data of M vs. H above $T_B$ is interpreted in terms of the particle size distribution, resulting in $\langle \mu \rangle$ consistent with calculated values. The effect of varying the core size on the magnetic properties in the Fe@Pt system will be undertaken in future studies.

**Acknowledgments**

This work was supported in part by NSF grant #DGE-1144676. We acknowledge use of WVU Shared Research Facilities.

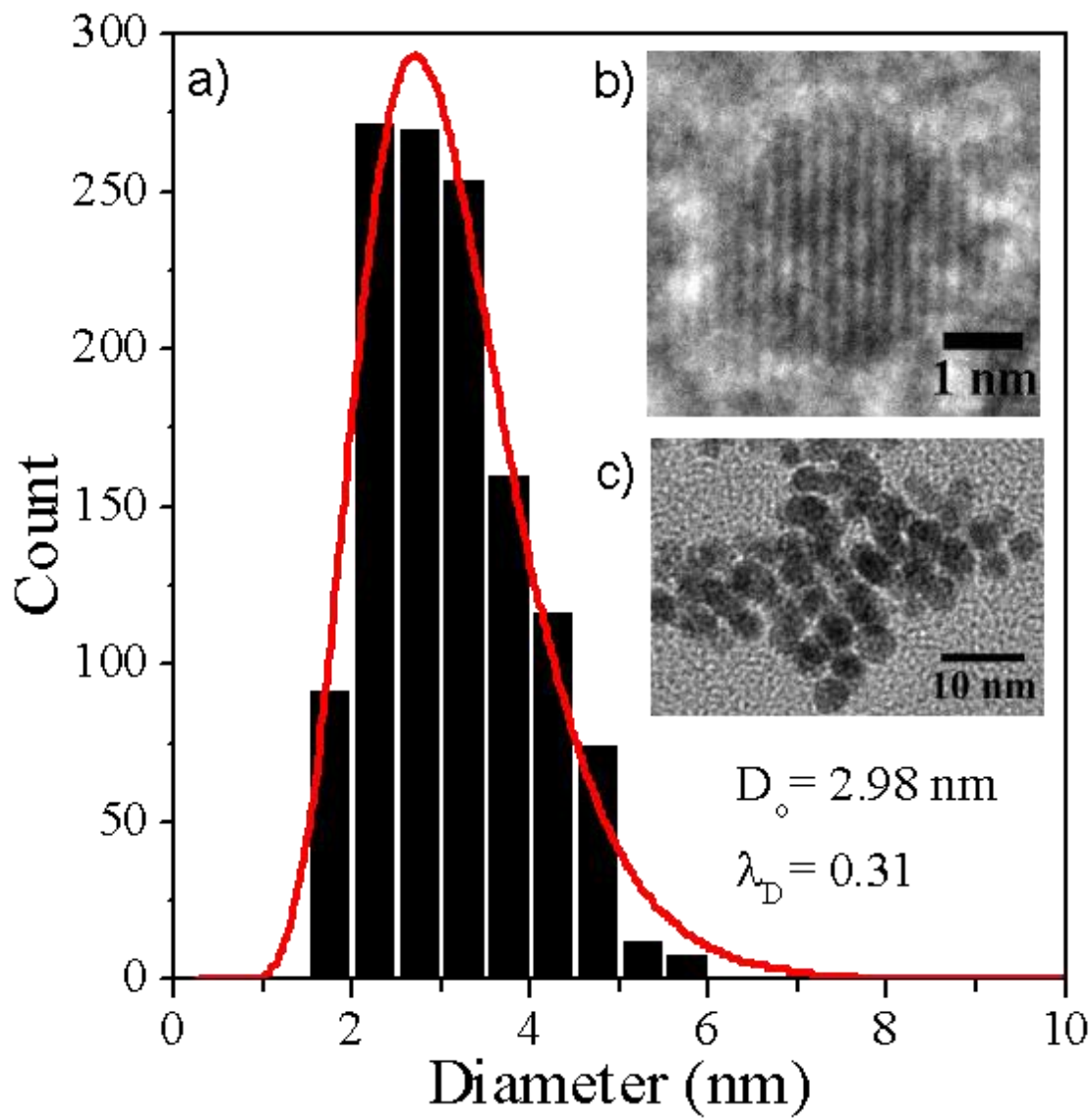

**Figure 1: a)** Histogram of measured particle diameters fitted to a lognormal distribution function (solid line) **b)** High resolution TEM image of one NP shows lattice spacing corresponding to Pt $d_{111}$ (scale bar is 1 nm) **c)** TEM image showing nearly spherical particles (scale bar is 10 nm)

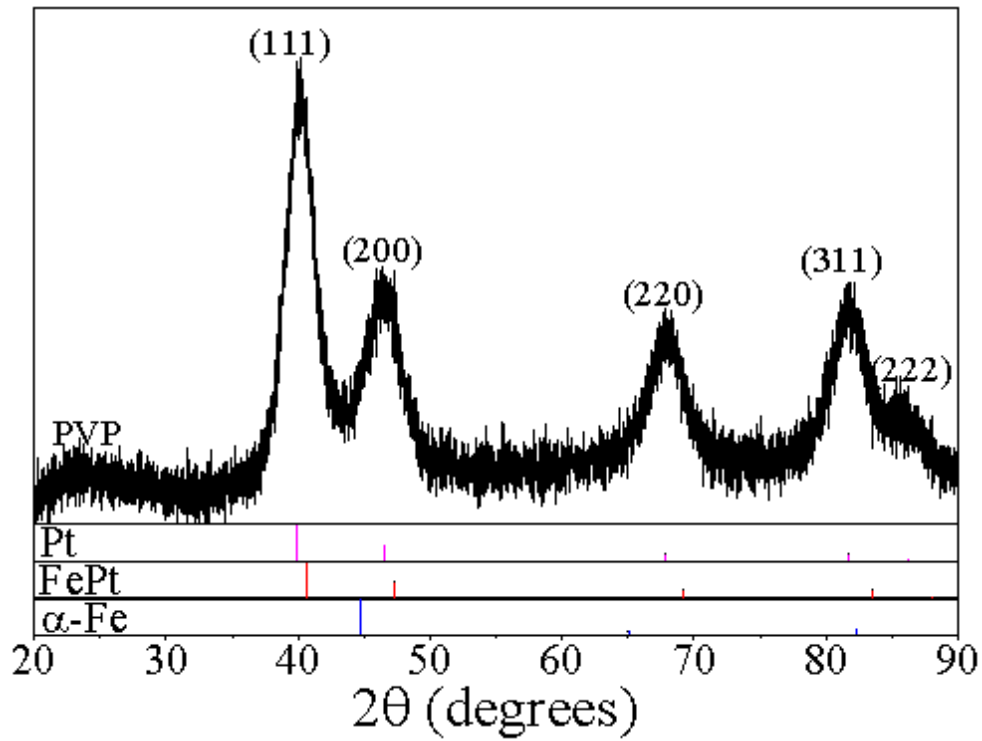

**Figure 2:** XRD pattern with peaks indexed for Pt. Below the pattern are the expected peak locations and relative intensities for Pt, cubic FePt, and α-Fe.

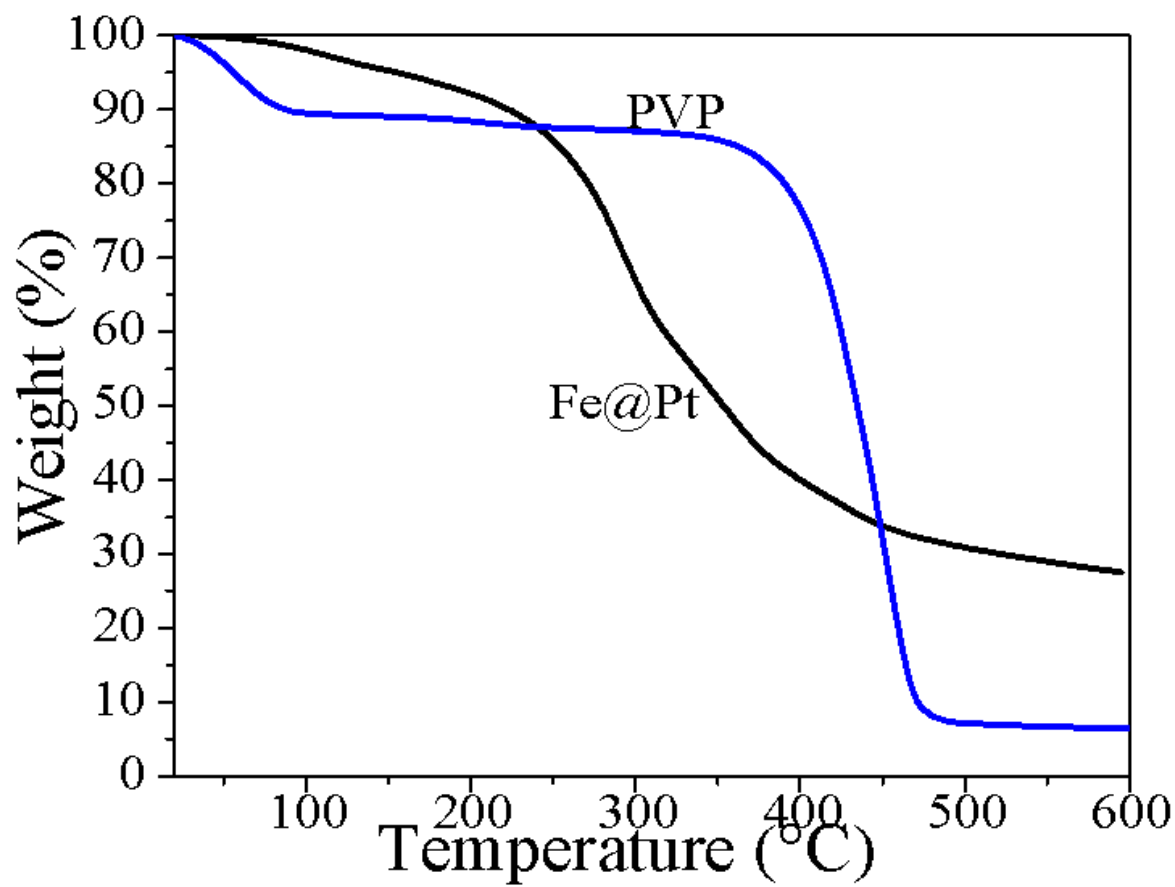

**Figure 3:** TGA data for pure PVP and the Fe@Pt NPs indicate that 73% of the measured sample mass comes from PVP present on the nanoparticles.

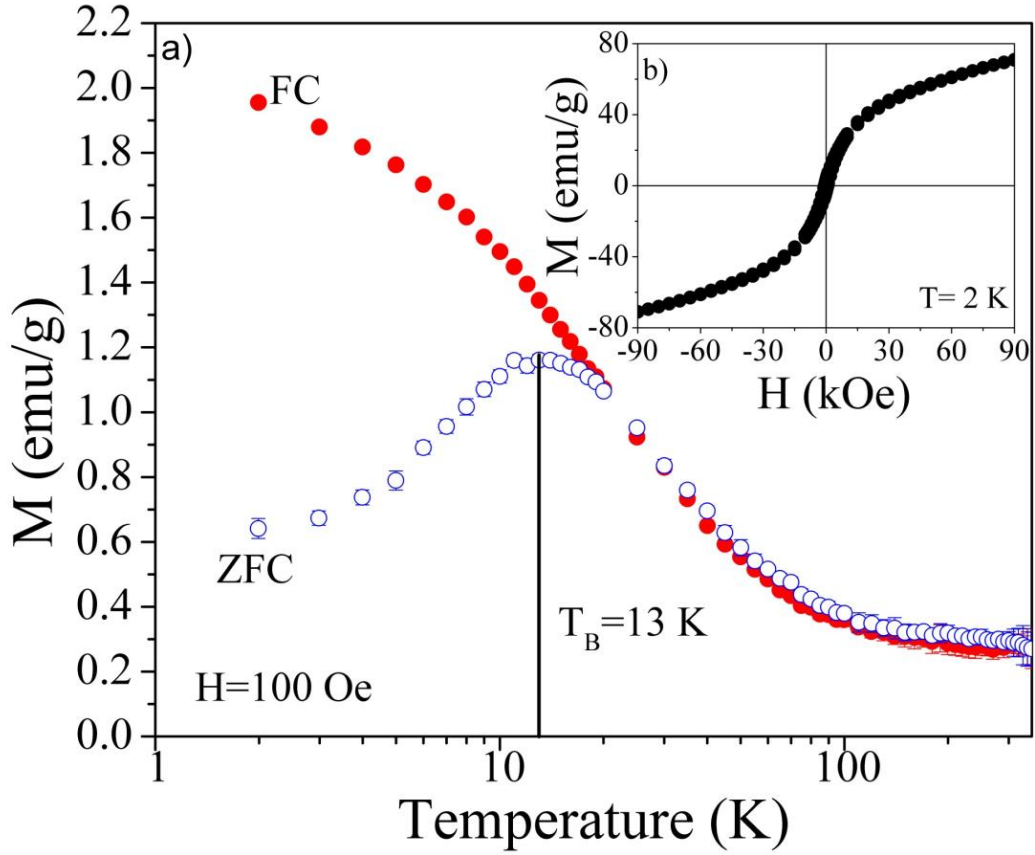

**Figure 4: a)** M vs. T for the ZFC (open circles) and FC (closed circles) cases **b)** Measured M vs. H hysteresis loop at T= 2 K (see Fig. 5 for a view of the low-field region).

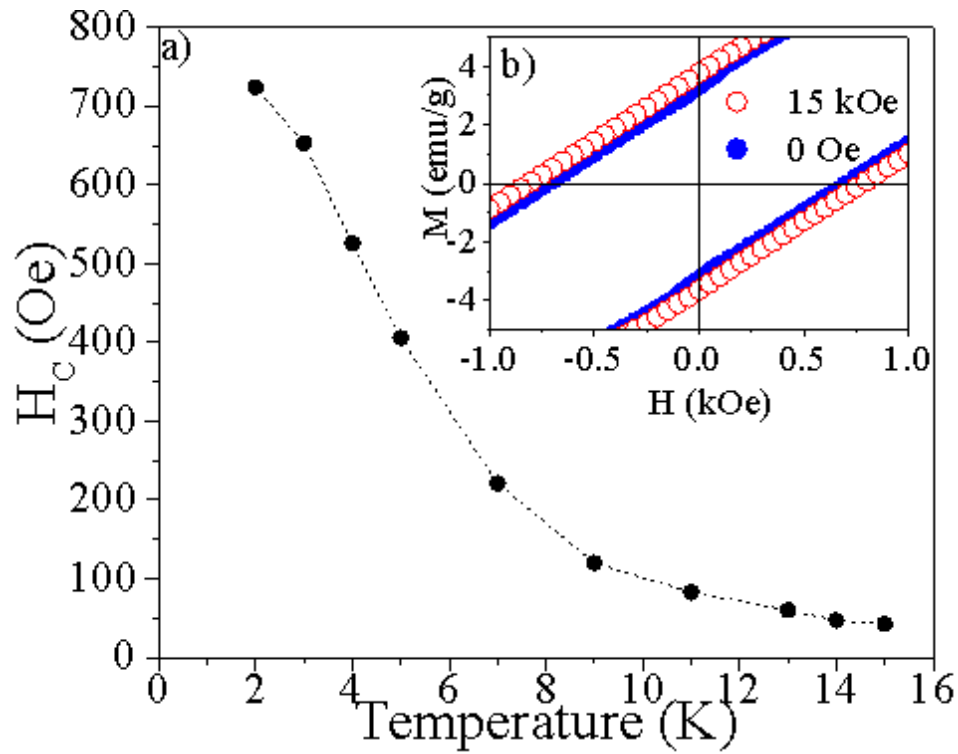

**Figure 5: a)** Measured values of coercivity $H_C$ vs. T for the ZFC case; **b)** Low field region of hysteresis curves for ZFC (filled circles) and FC (open circles) in 15 kOe.

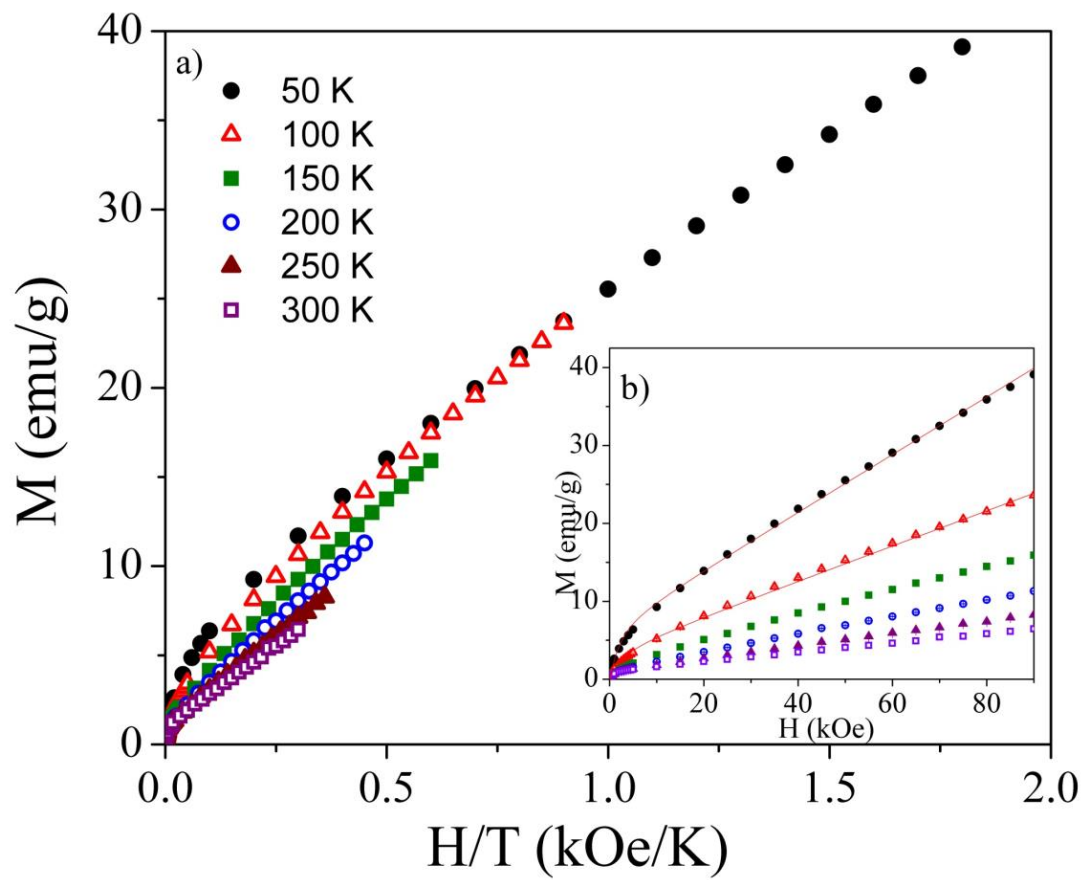

**Figure 6:** a) M vs H/T b) M vs H; the solid lines for the 50 K and 100 K data are fit to Eq. 3.

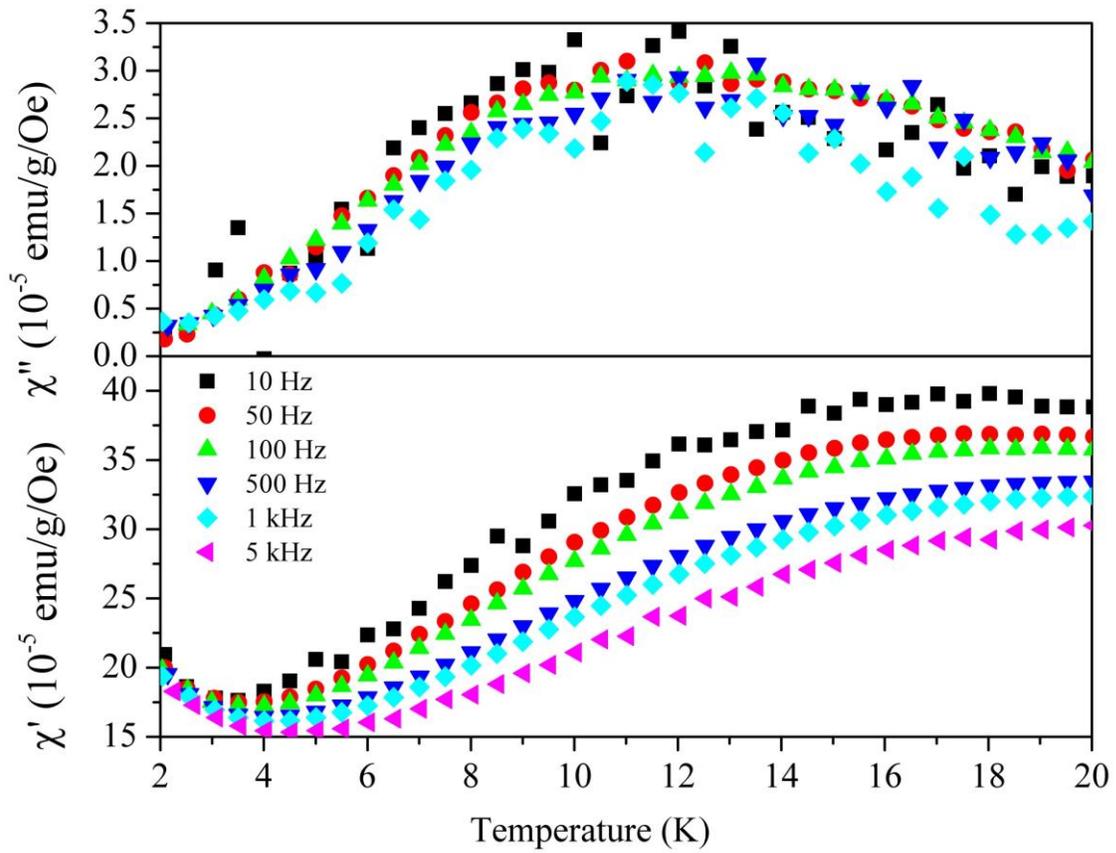

**Figure 7**: In-phase and out-of-phase ac susceptibilities vs temperature.

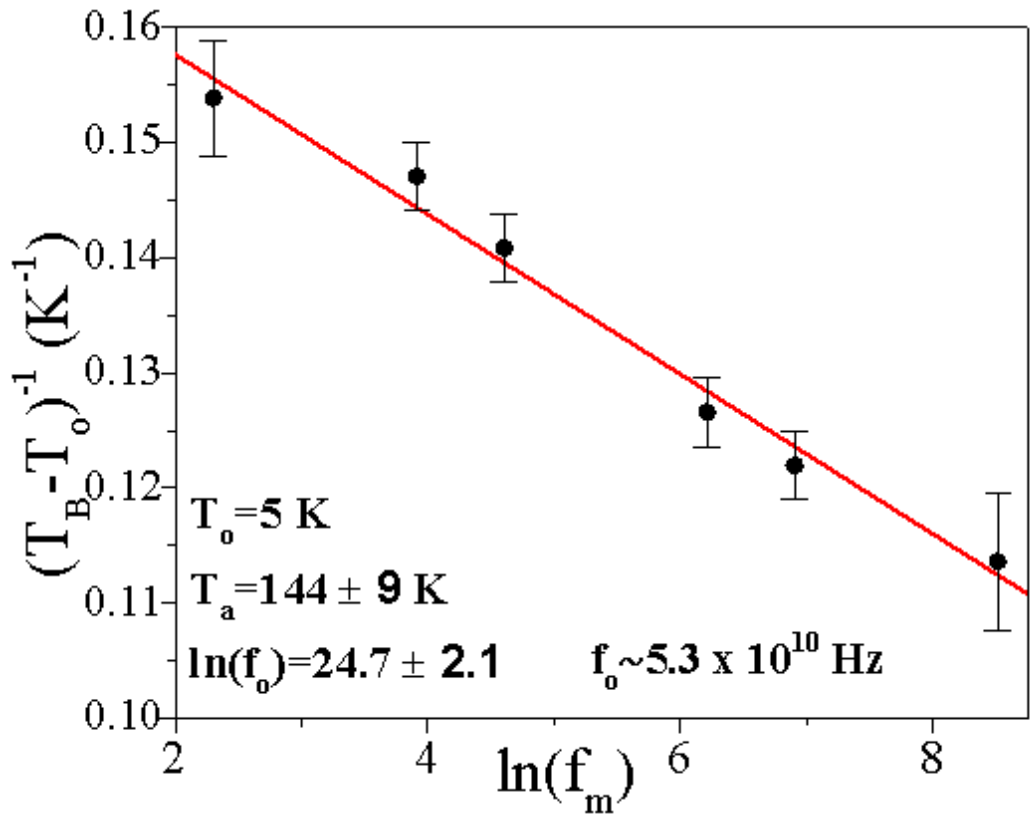

**Figure 8**: Plot of $(T_B-T_o)^{-1}$ vs $\ln(f_m)$ determined from $d(\chi'T)/dT$. The solid line is a fit to the data for $T_o= 5$ K.